\documentclass[aps, unsortedaddress,superscriptaddress,amsmath,amssymb,showpacs]{revtex4}
\usepackage{graphicx}
\usepackage{epsfig}
\include{graphics}
\begin{document}


\bibliographystyle{unsrt}

\sloppy

\newcommand{\ord}[1]{\mathcal{O}(#1)}
\newcommand{\rl}[1]{#1_{12}}
\newcommand{\eps}{\epsilon}


\title{A Liquid Film Motor}

\author{A. Amjadi}
\affiliation{Sharif University of Technology, Department of Physics,
P.O. Box 11365-9161, Tehran, Iran.}

\author{R. Shirsavar}
\email{shirsavar@physics.sharif.edu} \affiliation{Sharif University
of Technology, Department of Physics, P.O. Box 11365-9161, Tehran,
Iran.}

\author{N. Hamedani Radja}
\affiliation {Instituut-Lorentz, Universiteit Leiden, Postbus 9506, 2300 RA, Leiden, The Netherlands} \affiliation{Institute for
Studies in Theoretical Physics and Mathematics, P.O. Box 19395-5531,
Tehran, Iran.}

\author{M. R. Ejtehadi}
\email{ejtehadi@sharif.edu} \affiliation{Sharif University of
Technology, Department of Physics, P.O. Box 11365-9161, Tehran,
Iran.}

\date{\today}

\begin{abstract}
It is well known that electro-hydrodynamical effects in freely suspended liquid films can flow the liquid. Here we report a purely electrically driven rotation in water and
some other liquid suspended films with full control on the velocity and the chirality of the rotating vortices. The device, which is called ``film motor'', consists of a quasi two-dimensional electrolysis cell in an external in-plane electric field, crossing the mean electrolysis
current density. If either the external field or the electrolysis
voltage exceeds some threshold (while the other one is not zero),
the liquid film begins to rotate. The device works perfectly with both DC and AC fields.

\end{abstract}
\pacs{47.32.Ef, 68.15.+e, 47.57.jd}

\maketitle

\newpage
\section{Introduction}


\label{Introduction} 
In recent years, scientists have become
interested in the physics of liquid films. Macroscopic thin films
are important in physics, biophysics, and engineering. They can be
composed not only of complex materials such as polymer solutions,
but also of common liquids such as water or oil. When the films are
subjected to the action of various chemical, thermal, structural or
electrical factors, they display interesting dynamical phenomena
such as wave propagation, wave steepening, and chaotic
responses\cite{Steepening,Propagating}. Water films are more
mysterious as the physics of the hydrogen bonds in the water film
interface is not completely understood yet\cite{Nelson,Jones}.
Although, it is easy to produce pure water films, dissolving
surfactant molecules in water makes the films more stable and
thinner. Suspended liquid films as thin as hundreds of nanometers or
less let us study physical phenomena in a quasi-two-dimensional
media. For example, quasi-two-dimensional vortices in magnetically
active films have been studied as a case of 2D
turbulence~\cite{Conformal}.

The effect of the electric field on the liquid films has been
studied widely, both in confined films between two transparent
plates~\cite{LiquidCrystals1, LiquidCrystal2}, and freely
non-confined suspended films~\cite{Suspendedfilms}. In the latter
case the field can produce electro-hydrodynamical (EHD) flow in the
films. The two-dimensional EHD motion was observed in freely
suspended films made of certain thermotropic liquid crystal phases:
nematic~\cite{static,Electrohydrodynamic} and smectic
A~\cite{Electroconvection1,Electroconvection2}. By applying a
sufficiently large electric field on thin film of nematic or smectic
liquid crystals a pattern of convective vortices is observed. When
the electric voltage is increased up to a threshold, some vortices
are formed on the film. The vortices rotate in two directions, and
the velocity of the rotations depends on the magnitude of the
electric voltage. The threshold voltage to enter to the vortex mode
is proportional to both average film thickness and square root of
electric conductivity of the film~\cite{Suspendedfilms}. The
presence of excess charges at the film surfaces also plays a crucial
role in formation of the vortex pattern.

The EHD motions also have been reported on the films of aqueous
solutions~\cite{Electrokenitics}. The driven motions is used in
micro-fluidic systems to mix or phase separate liquids or particles
in micro scale systems~\cite{microfluidics}.

Here, we introduce an EHD effect on freely suspended liquid films.
The effect can produce purely electrically induced rotations on
suspended films of water and some other polar liquids.  We call the
device ``liquid film motor", as the direction and the speed of the
rotation are controlled through the direction and strength of the
applied electric fields. The motor works perfectly just with pure
water, but to increase the film's life-time, we can dissolve some
amount of glycerin and a little detergent in the water.  In this
way, we make more stable films with micro- to nano-scale thicknesses
that rotate up to several minutes before they break.

\section{The experiment}
\label{experiment}
\subsection{The set-up}

The device consists of a 2D frame with two electrodes on the sides
for electrolysis of water films. (Fig.~\ref{fig1}). The frame is
made of an ordinary blank printed circuit board with a rectangular
hole at the center, and two copper strips on the sides of the
rectangular hole as electrodes. Immersing the frame into a liquid
and bringing it out simply makes a suspended liquid film on the
frame. As water wets copper perfectly, connecting the electrodes to
electrolysis voltage $V_{\rm el}$ causes average density of electric
current $\bf{J}_{\rm el}$ in the liquid. The cell is located between
two plates of a large capacitor. The capacitor produces an external
electric field $\bf{E}_{\rm ext}$ in the plane of the film and
almost perpendicular to the mean current density.

\subsection{The rotation}
\begin{figure}
\center
\includegraphics[scale=0.4]{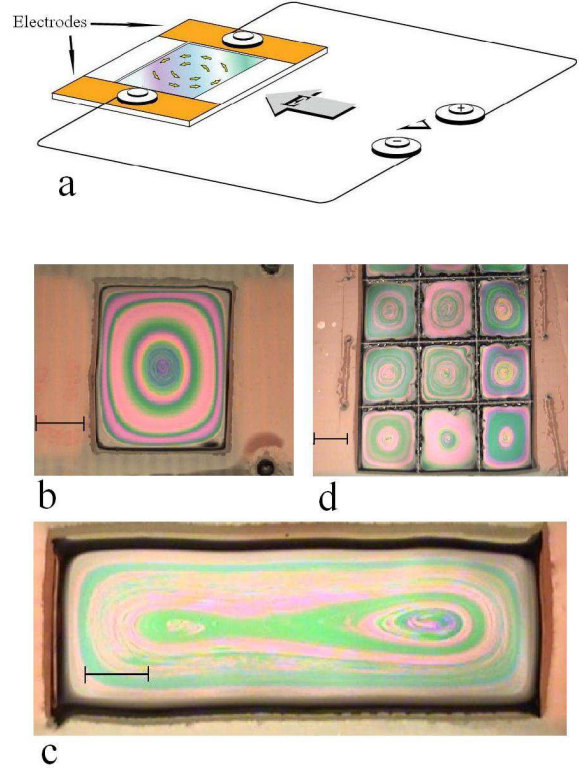}
\caption{\label{fig1}(a) The set-up of the experiment. (b) A
rotating soap film. Different colors on the film are the effect of
local thicknesses. (c) Because of the effect of boundary condition,
the pattern of rotation may be affected by the shape of the frame. A
rectangular cell produces two vortices with the same direction of
rotations. (d) The film is meshed by cotton strings. The direction
of the rotations in all compartments are the same. Bars on the
figures indicate scales of one centimeter. (Movies are available on
http://softmatter.cscm.ir/FilmMotor)}
\end{figure}

If either the external field or the electrolysis voltage exceeds
some thresholds (while the other one is not zero), the liquid film
begins to rotate (Fig.~\ref{fig1}). The direction and the speed of
the rotation can be controlled through the direction and strength of
the current and/or external electric field. The motor works
perfectly with just pure water, but to increase the film's
life-time, we can dissolve some amount of glycerin and a little
detergent in the water. Depending on the aspect ratio of the film
frame, the applied external electric field may produce several
vortices with different radii (Fig.~\ref{fig1}b and ~\ref{fig1}c),
but all of the vortices rotate in the same direction. This shows
that the vortices are not convective vortices. Increasing either
$V_{\rm el}$ or $E_{\rm ext}$ increases the vortex rotation speed.
To measure the rotation speed a digital video camera which takes 24
frames per second is employed. The color pattern of the film,
because of white light interference on the film, can be followed on
successive frames. In this way we are also able to find angular
velocity profile of the film.

\begin{figure}
\center
\includegraphics[scale=0.34]{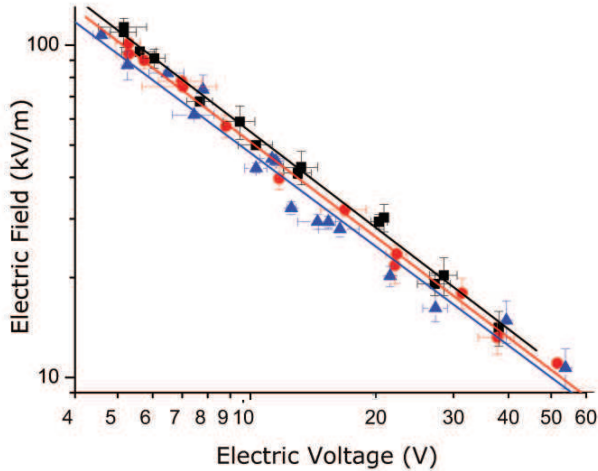}
\caption{\label{fig2}The electric field versus the electric voltage
at the rotating threshold for the solutions with different volume
fractions of glycerin ($c$). $c=0.1$ (Square). $c=0.3$ (Circle).
$c=0.5$ (Triangle).}
\end{figure}

The experiment shows that the threshold values of the fields to
start rotation in the film are highly related to each other. To see
this we fix the value of one of the fields and increase the other
until the film start rotation. The measured external-field
thresholds for starting the rotations at different electrolysis
voltages, are shown in Fig.~\ref{fig2}. The log-log plot of the
electric field versus electrolysis voltage indicates a slope of
$-1.00 \pm 0.02$ for solutions with different glycerin
concentration, $C$. Thus the experiments propose that the threshold
fields obey simple scaling relation of $V_{\rm el} E_{\rm ext} =
{\rm Const}$. The constant value depends on the average and the
profile of the film thickness. To fix this parameters we tilt the
water film plane by $\theta=5^{\circ}$ from the horizon. In this way
the thickness profile changes continuously with time and let us to
manage the experiment when it reaches to the similar patterns of
interference.

\subsection{The direction of the rotation}
The direction of the rotation of the film is on the direction of
$\bf{E}_{\rm ext}\times \bf{J}_{\rm el}$. If $\bf{E}_{\rm ext}$ is
perpendicular to $\bf{J}_{\rm el}$, we have the maximum rotation
speed. While by decreasing the angle between $\bf{E}_{\rm ext}$ and
$\bf{J}_{\rm el}$ from $90 ^{\circ}$ to $0 ^{\circ}$, the rotation
velocity monotonically drops down to zero.

Depends on the shape of the cell it is possible that one observes
more than one vortex on the film. For example, in the case of the
rectangular frames with large enough aspect ratio of the sides two
vortices with the same direction of rotation are observed. Also if
we divide the cell to smaller compartments using cotton strings
there are individual vortices in separate compartments all of
rotating in the same direction obeying the above mentioned rule.

\section{The rotation mechanism }
\label{mechanism}

There are several charge based mechanisms which can be responsible
as the source of the rotations. For example non-uniform ion
distribution, caused by the external electric field (the
electrophoresis effect), can break the symmetry under translation of
the frictional forces acting on the liquid molecules. An other
mechanism has been suggested by Chiragwandi {\it et al.} to explain
observed micro-scale vortices in water based
transistors~\cite{vortex}. Their mechanism is based on ionization of
the water molecules in the electrolysis double layers close to the
electrodes.

By applying the boundary condition on the electric field components
excess charges on the liquid-air interface exist. To explain the
electrically induced convective vortices, which has been observed
experimentally in suspended films of liquid crystals by Morris and
his coworkers for smectic~$A$
~\cite{Electroconvection1,Electroconvection2}, and by Faetti {\it et
al.} for Nematic ~\cite{static,Electrohydrodynamic} this surface
charges are employed.

\begin{figure}
\center
\includegraphics[scale=0.34]{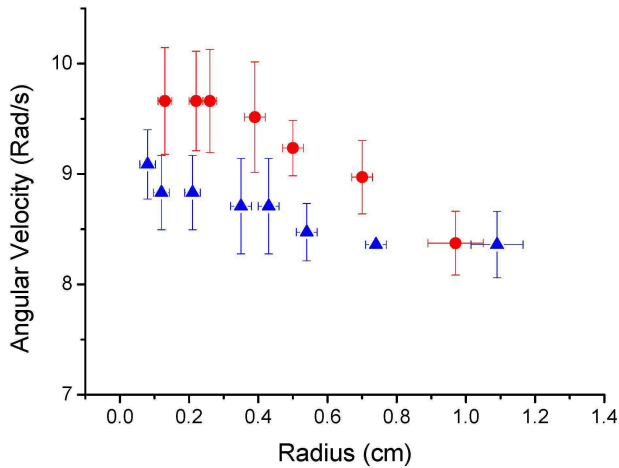}
\caption{\label{fig3} The film's angular velocity decreases
monotonically with the radial distance from the vortex center. We
have waited for $45$ seconds (circles) and $7$ minutes (triangles)
after the rotation has started.}
\end{figure}

However all above mentioned charge based mechanisms could be
relevant to the rotation and at least they predict the direction of
rotation as observed, we have enough reasons to show non of them is
dominant mechanism. In all of them the source of rotation is sitting
on the borders of the cell. But investigating angular velocity of
vortex shows it rotates faster in the center (fig~\ref{fig3}).  Also
changing the conductivity of the water by adding salt to it do not
change speed of rotation significantly while it can increase the
conductivity by orders of magnitude. There are also other
observations which can not be explained by these suggested
mechanism.  When we divide the electrolysis cell to a number of
separate compartments using non-conducting cotton strings
(Fig~\ref{fig1}d),it is interesting that we observe similar
rotations in different compartments. looking at the compartment in
the middle, surely it has different charge distribution both in
volume and surface and there is no ionization chemical reaction on
its boundaries, but it rotates as fast as the outer compartments.

On the other hand we have observed that in addition to water films,
the film motor operate with some other polar liquids, e.g. Aniline,
Anisole, Chlorobenzene and Diethyleoxalate. The characteristics of
the rotation for these liquids are similar to each other, then the
mechanism of the rotation could be the same for all of them. For
example, the direction of the rotation of all liquids that rotate,
obeys the same previously mentioned simple rule. Also the threshold
values of the fields are in the same order of magnitudes for all the
liquids with different electrical conductivity, viscosity and/or
density.

We do not observe a clear rotation in the film of non-polar liquids.
In particular for 1-Dodecene which is a non-polar liquid with stable
films, we do not see any induced rotation. This suggests that the
intrinsic polarity of the liquid molecules should involve in the
mechanism of the rotation. According to these experiments any long
lasting film of liquids with polar molecules rotate well, even if
they have a very small conductivity and/or do not contain hydrogen
bonds.

\section{The rotation with AC electric fields}
\label{AC}
\subsection{The rotation without electro-chemical reactions}

To eliminate any electro-chemical reactions near two electrodes that
are connected to a liquid film, we have used electrodes covered with
insulating paint. Passing the electric current through the liquid
film in this case is possible by applying an alternating voltage.
With this technique, the electric current passes without any
chemical reactions near the electrodes.

If we repeat the experiment with alternating electric voltage and
current (in phase with same frequency), the film will rotate. Our
experiments on water films have been done with frequencies up to
$40$~kHz, and the characteristics of the rotation remain the same as
in DC case. Thus, the mechanism for the rotation is independent of
the electro-chemical reactions.

\subsection{The phase and frequency effect on the rotation}

Applying the external and internal fields on a liquid film, with
different frequencies, does not produce rotating vortices, but it
causes vibrational movements. If the applied alternating fields have
exactly the same frequencies, the liquid film rotates. In this case,
the threshold and the velocity of the rotation, depend on the phase
difference between the fields, $\phi$ as well as the magnitude of
the fields.

\begin{figure}
\center
\includegraphics[scale=0.34]{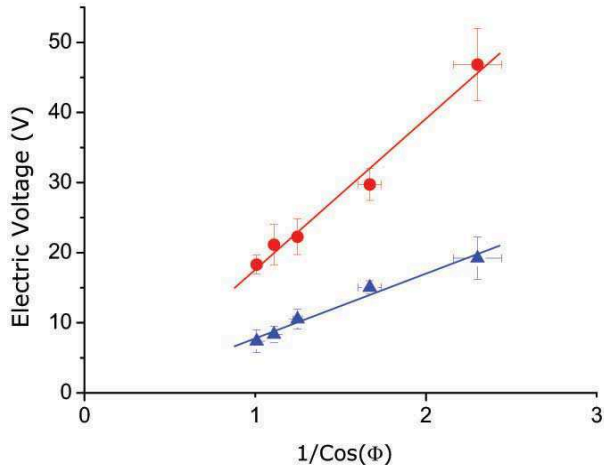}
\caption{\label{fig4} Threshold voltage, $V_{0}$, versus
$1/\texttt{Cos}(\phi)$ for two different values of $E_{0}$.
$E_{0}=50$ $kV/m$ (circle) and $E_{0}=100$ $kV/m$ (triangle). In
both experiments, the fields frequency are 50 Hz.}
\end{figure}

For investigation of the effect of phase difference on the
threshold, we have applied alternating electric field and voltage,
\begin{equation}
\label{Eq:1} E(t)= E_0 \texttt{Sin}(2\pi ft)
\end{equation}
and

\begin{equation}
\label{Eq:2}  V(t)= V_0 \texttt{Sin}(2\pi ft+\phi),
\end{equation}
with $f=50$~Hz. By changing the magnitude of the phase difference,
the voltage threshold changes while the magnitude of external
electric field is kept constant. For two different given $E_0$, the
values of $V_0$ in terms of $\cos\phi$ is shown at threshold
points(Fig.~\ref{fig4}). To justify these experimental data, we
suppose that at threshold points, the time average of $EV$,
$\overline{EV}$, should be constant. Having
\begin{equation}
\label{eq:eq3} \overline{EV}= E_{\texttt{rms}}V_{\texttt{rms}}
\texttt{Cos}(\phi),
\end{equation}
thus,
\begin{equation}
\label{eq:eq4}E_{\texttt{rms}}V_{\texttt{rms}}\propto
\texttt{Cos}(\phi)^{-1}.
\end{equation}
This argument is well consistent with experimental data
(Fig.~\ref{fig4}). If the frequencies of the fields are not the same
the time average in large intervals is zero and the film only
vibrates. The vibration frequency is the beating frequency.

\section{Conclusions}
\label{conclusions}

Our experimental observations have shown that the suspended films of
some liquids can rotate by applying pure electric fields. In
contrast with the convective vortices in suspended films, speed an
direction of the rotation in the case of our film motor is fully
under control by the mean values of the crossed electric fields. In
our experiments the films with length scales from tenth of
millimeter up to several centimeters have been examined, and there
is no reason that it does not work in smaller length scales. This
phenomenon may have wide industrial application, e.g. in liquid
based centrifuge or liquid mixing devices.

Liquid films rotate without any electro-chemical reactions near the
electrodes. Dissolving some amount of a salt in a pure liquid,
although increases the electrical conductivity by a few orders of
magnitude, but has not a notable effect on the rotation velocity and
threshold. Therefore, the ion movement have not significant effect
in the rotation, in contrast to the role of the intrinsic dipole
moment of a liquid molecules which is deeply vital. Examining
different liquids shows that polar liquids rotate well.

Any efforts to rotate a bulk of liquid was defeated. The fact that
only thin liquid films rotate notably and that rotation can not be
observed in relatively thick films even at high fields, implies that
this phenomenon is a surface effect.

{\bf Acknowledgment:} We are indebted to S. W. Morris and K. Malek
for their critical comments and also to F. Gobal, A. Nejati, M.
Mosayebi, M. Sahimi, S. Rahvar for their invaluable help. M.R.E is
thankful for the partial support of the Center of excellence in
Complex systems and Condensed matter physics (CSCM). We also are
thankful for partial support of the Sharif applied physics research
center. All experiments have been done at Laser and Medical Physics
Lab of Sharif University of Technology.

\end{document}